# Remote detection of explosives using Field Asymmetric Ion Mobility Spectrometer installed on multicopter

*AUTHOR NAMES*

Yury Kostyukevich[a,b,c,d], Denis Efremov[e] Vladimir Ionov[f] Eugene Kukaev[d] and Eugene Nikolaev[a,b,c,d]

AUTHOR ADDRESS

[a] Skolkovo Institute of Science and Technology Novaya St., 100, Skolkovo 143025 Russian Federation

[b] Institute for Energy Problems of Chemical Physics Russian Academy of Sciences Leninskij pr. 38 k.2, 119334 Moscow, Russia;

[c] Emanuel Institute for Biochemical Physics Russian Academy of Sciences Kosygina st. 4, 119334 Moscow, Russia.

[d] Moscow Institute of Physics and Technology, 141700 Dolgoprudnyi, Moscow Region, Russia

[e] Private educational institution of higher professional education Moscow Technological Institute Leninskii prospect 38a, Moscow, 119334

[f] "Lavanda-U Limited" 111123, Moscow, shosse Entuziastov, d. 56, p. 27

ABSTRACT

The detection of explosives and drugs in hard-to-reach places is a considerable challenge. We report the development and initial experimental characterization of the air analysis system that includes Field Asymmetric Ion Mobility Spectrometer (FAIMS), array of the semiconductor gas sensors and is installed on multicopter. The system was developed based on the commercially available DJI Matrix 100 platform. For data collection and communication with operator, the special compact computer (Intel Compute Stick) was installed onboard. The total weight of the system was 3.3 kg. The system allows the 15 minutes flight and provide the remote access to the obtained data. The developed system can be effectively used for the detection of impurities in the air, ecology monitoring, detection of chemical warfare agents and explosives, what is especially important in light of recent terroristic attacks. The capabilities of the system was tested on the several explosives such as trinitrotoluene and nitro powder.

TEXT

Mass spectrometry is one of the most sensitive and accurate analytical techniques for the investigation of the chemical composition of the unknown substance and has found various application in the chemistry, forensic science, food and drug monitoring etc. The only disadvantage of the mass spectrometer that limits filed use is the weight and power consumption.

However, there are many cases when the sample collection and its transfer to the laboratory is impossible. For example, the unknown chemical weapon was recently used in Syrian city Khan Sheikhoun. This substance could be easily identified by a compact mass spectrometer carried by a drone. At the same time, sending troops to collect the sample would inevitably include the considerable risk.

The first installation of the mass spectrometer on the aircraft dates back to the 50-th, when Bennett radio frequency mass spectrometers where installed on the sounding rockets and artificial satellites and were used for the investigation of the ionic composition of the upper atmosphere[1-3]. Further development of the field resulted in the introduction of more complicated aircraft-borne mass spectrometers designed for specific purposes: aerosol mass spectrometers[4], dust impact mass analyzer[5] etc. Currently, all modern scientific spacecraft, such as Phoenix[6], Rosetta[7,8], Curiosity[9,10] are equipped with several mass spectrometers. Unfortunately, only few laboratories can afford the development of the aircraft-borne mass spectrometer and such instruments have considerable weight and size, what limits its use in the city.

Recently a considerable progress has been achieved in the developing of low weight and portable mass spectrometers[11-14]. Miniature ion traps have found application in ecology monitoring[15,16], detection of explosives[17] and drugs[18]. Miniature unmanned aircraft-borne mass spectrometer was used for the in-situ volcanic plume analysis[19]. Considerable successes have been made in the area of miniaturized and chip-based FAIMS[20-22]. Portable hand-held Ion mobility spectrometers (IMS) have been used for dietary supplements screening[23], detection of drugs, explosives and chemical warfare[24,25].

In the present paper we describe our approach for the developing portable, low weight aircraft-born measurement system that includes FAIMS and the array of gas sensors. The total weight of whole system is 3.3 kg, and the dimensions are 52*52*22 cm. The low cost and compact size allow our system being potentially suitable for the ecology monitoring in the cities and for the detection of chemical warfare agents.

METHODS

**The flying vehicle.** We have chosen a DJI matrice-100 as an aircraft to carry the measuring system. This multirotor unmanned aircraft vehicle (UAV) includes the frame, adapted for the mounting of additional custom equipment; the motor assembly able to keep the UAV in the air to 40 minutes without payload; flight controller with GPS positioning with accuracy of hovering up to 2 meters; the control system over the distance of up to 2 km; the onboard telemetry data transmission system.

**Gas sensors.** The array of the semiconductor gas sensors (DFRobot company, Shanghai, China ) was installed on the UAV. Semiconductor gas sensors measures the changes of the conductivity of the thing film after the adsorption of the target gas[26]. Such devises are compact nave low energy consumption and do not require pumping. The array of gas sensors sensible to different gases can be used as a complementary to FAIMS technique. The detailed description of the sensors is given in the Table 1.

**FAIMS spectrometer.** The FAIMS spectrometer was developed by the company Lavanda-U Limited and was initially optimized for the detection of the explosives. The schematic design is presented in the Fig.1. The FAIMS spectrometer consists of the ion source ($^{63}$Ni), 2 coaxial electrodes, detection plates and the flow driver. The gas flow rate was equal to 6 l/min. The asymmetric impulse voltage with repeating frequency 100 kHz, duty cycle 3:1 and amplitude 5 kV was used as a separating voltage. The FAIMS spectrometer requires 12V; 500mA power supply what enable its feeding from main copter battery.

**Data acquisition and transfer system.** The Intel Compute Stick with installed Windows 10 was used as a main onboard computer. This device has the weight of 80 g and requires only 5V power supply what enable its feeding from main copter battery. The data from the gas sensors were collected by the Arduino UNO microcontroller and transferred to the onboard computer. To enable the communication with the operator the remote desktop access to the onboard computer was provided using TeamViewer 12 software. To enable the internet access, the 4G modem was installed onboard (we used Yota LTE 4G).

RESULTS AND DISCUSSION

The photo of the assembled system is presented in the Fig.2. The total weight of the system was equal to 3.3 kg what is lower than the maximum weight that can carry the DJI matrice 100 (3.5 kg). The array of gas sensors was installed on top of the DJI matrice 100 platform (Fig.2 A). The onboard computer and FAIMS spectrometer were installed below the platform (Fig.2 C,D). The photo of the hovering system is presented in Fig.2B.

It was observed, that the developed system provided the stable data acquisition and transmission. The readings of gas sensors and FAIMS spectrometer were transmitted to the smartphone fixed on the remote control. This allowed the operator to simultaneously control the flight and read the information from the sensors. All data were saved to file and additionally the video from smartphone screen was recorded.

First we have investigated the capability of the developed system to detect the volatile organic in the air. For this experiment the FAIMS tube was used without the sucking cup (Fig.2D). To create the cloud of target gas we have used the spray can with the perfume. After several injections were made in the air, the quadrocopter flew through this region and data were acquired. We have observed that the major technical obstacle for performing the measurements was the powerful airflow created by the rotating propellers. The wind was so strong that it instantly mixed the cloud of gas with the surrounding atmosphere what resulted in the rapid decrease of the concentration of the measured gas. As a consequence, the developed system can be used for the performing measurements in the homogeneous atmosphere, but considerable improvement is required for performing the measurements in the local regions such as: exhaust of ventilation pipes, vicinity of the small suspicious objects etc. Although, the system is suitable for the ecology monitoring and detection of chemical warfare agents.

The results of the measurements are presented in the Fig.3. It can be seen that when the quadrocopter hovers in the clean atmosphere the reading of all gas sensors remains stable, but when the copter fly through the cloud of odorous gas, all gas sensors (except for $H_2$ and $CO_2$ sensors) responded. The response of almost all sensors is expectable, because the sensors are not selective and differ only by the sensitivity to particular substances (see Fig.3 A,B). One of the disadvantages of the semiconductor gas sensors before the FAIMS is the considerable time required for the attenuation of the signal. This time is determined by the desorption of the gas from the surface of the sensor. The FAIMS spectrometer doesn't have this disadvantage.

The ionograms recorded by the FAIMS spectrometer are presented in Fig.3C – the case of clean atmosphere and Fig.3D – the case when copter flew through the cloud of gas. The distinct differences between two ionograms can be clearly seen. The appearance of new peaks indicate the transmission of the ionized impurities through the separating electrodes assembly for certain compensation voltage. Though the FAIMS spectrometer cannot identify the unknown substance, it can be carefully tuned for the detection of the target gas.

After the validation of the system we have investigated its capability for the detection of the explosive. We have recorded the ionograms corresponding to the trace vapors of different chemical substances: trinitrotoluene (TNT), ionol, nitro powder and nitroglycerine. The results are presented in the Fig.4A. It can be clearly seen, that ionol is detected at the compensation voltage ~ 5.3V, TNT at ~8.5V and nitro powder and nitroglycerine at ~11V. To determine the limits of detection, we have recorded signals, corresponding to different concentration of TNT in the air. The ionograms are presented in the Fig.1S. The values of signal and signal/noise ratio are shown in the Fig.4B. We have estimated that the limit of detection for TNT is ~$1*10^{-13}$ $g/cm^3$.

After proving the capabilities of the system for detection of explosives we have performed the in-filed experiments. The design of the experiment is presented in the Fig.5. The traces of different explosives (TNT and nitroglycerine) were put on the ground at some distance from the operator. The operator flew quadrocopter to the places of the deposition of explosives. The quadrocopter landed and vapors of the explosives were sucked via a suction cup into the FAIMS. The suction cup was placed near the ground to increase the sensitivity. The results are presented in the Fig.5B. We have observed that it is possible to detect traces of explosives and even separate TNT and nitroglycerine. So, the remote detection of explosives in hard-to-reach place was performed.

CONCLUSION

We have developed a multicopter-based air monitoring system that utilize array of the gas sensors and FAIMS spectrometer. The developed system allows the detecting of air impurities in hard-to-reach places and transferring data online to the operator. The major disadvantage of the developed system is the powerful airflow created by the rotating propellers that mixes and homogenizes the atmosphere and, therefore, decreases the concentration of the target gas in the vicinity of the multicopter. Also, the turbulent wind can disturb the air flow inside the FAIMS tube. This limits the application of the developed system for the analysis of small objects, such as ventilation exhausts, but the system still can be used for the ecology monitoring and detection of chemical warfare agents. The total cost of the system includes multicopter (~5000$), FAIMS spectrometer (~3000$), onboard computer and 3G modem (~700$), array of gas sensors (~100$). The capabilities of the system was tested on the several explosives such as trinitrotoluene and nitro powder. The limit of detection for TNT is $\sim 1*10^{-13}$ g/cm$^3$, the system can separate the TNT and nitroglycerin. Therefore, the developed system is relatively low cost, what makes it possible for almost any laboratory to assemble the remote FAIMS spectrometer.

AUTHOR INFORMATION

Corresponding Author

* Eugene Nikolaev ennikolaev@rambler.ru

AUTHOR CONTRIBUTIONS

The manuscript was written through contributions of all authors. All authors have given approval to the final version of the manuscript

FUNDING SOURCES

The research was supported by the Russian Scientific Foundation grant № 14-24-00114.

COMPETING FINANCIAL INTEREST

Authors declare no competing financial interests.

TABLES

Table 1. The description of the used gas sensors.

| N | Sensor | Description |
|---|--------|-------------|
| 1 | MG811 | Electrochemical CO2 sensor |
| 2 | MQ9 | Semiconductor CO/Combustible Gas sensor |
| 3 | MQ8 | Semiconductor hydrogen gas sensor. |
| 4 | MQ7 | Carbon Monoxide Sensor. Concentrations in the air from 20 to 2000ppm. |
| 5 | MQ6 | Semiconductor Propane Gas Sensor. Can be used to detect iso-butane, propane and liquefied petroleum gas (LPG). |
| 6 | MQ5 | Semiconductor gas sensor, which is very sensitive with LPG, natural gas , town gas. It is less sensitive with alcohol, cooking fumes and cigarette smoke |
| 7 | MQ4 | Semiconductor gas sensor, which is able to detect nature gas, CH4. It is less sensitivity in detecting alcohol, cooking fumes, and cigarette smoke. |

| 8 | MQ3 | Semiconductor alcohol sensor. It has a high sensitivity to alcohol and less sensitivity to benzine. |
| 9 | MQ2 | Semiconductor gas sensor, which is able to detect LPG, i-butane, propane, methane ,alcohol, hydrogen, smoke. |

FIGURES

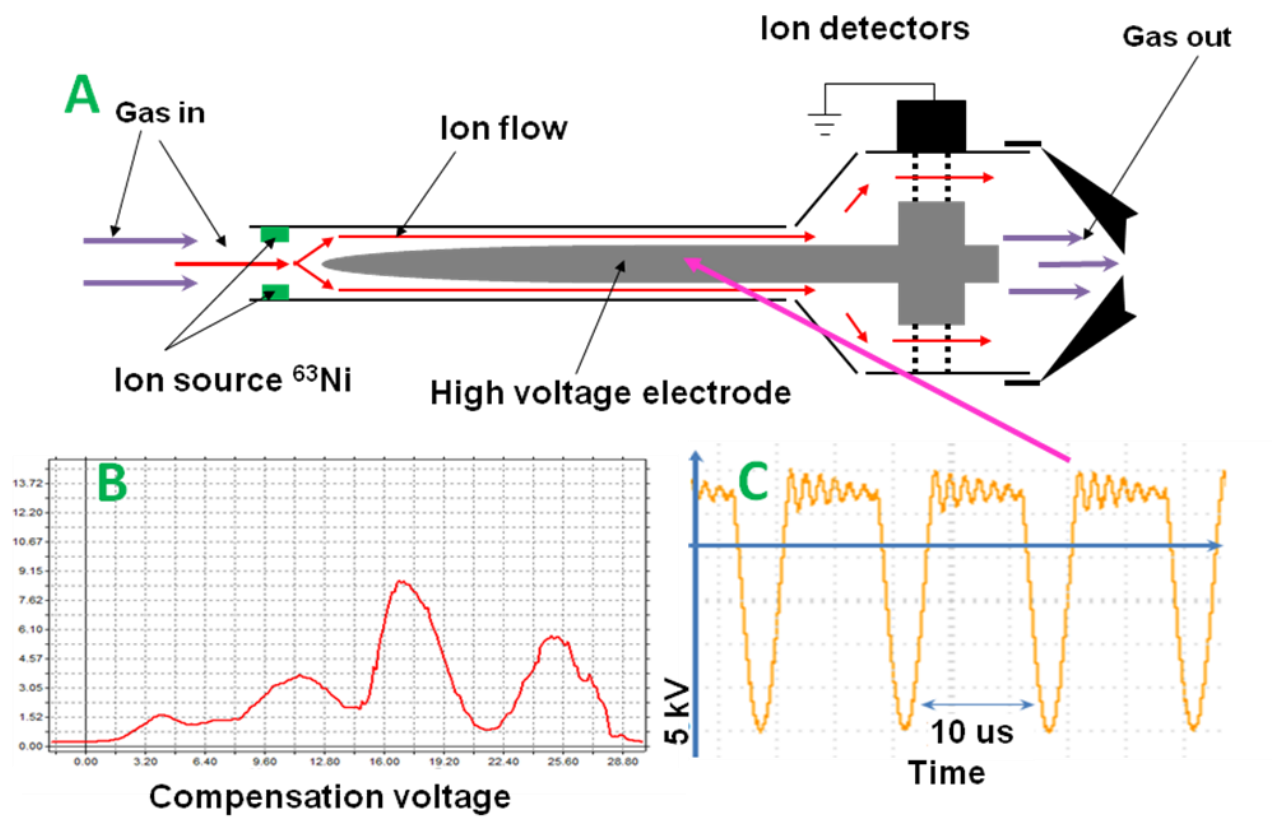

Fig.1 The schematic design of the FAIMS spectrometer developed by Lavanda-U Ltd. that was used in the present study. A – the schematic design of the device, B – the dependence of the signal on the compensation voltage. C – the shape of the potential applied to the high voltage electrode.

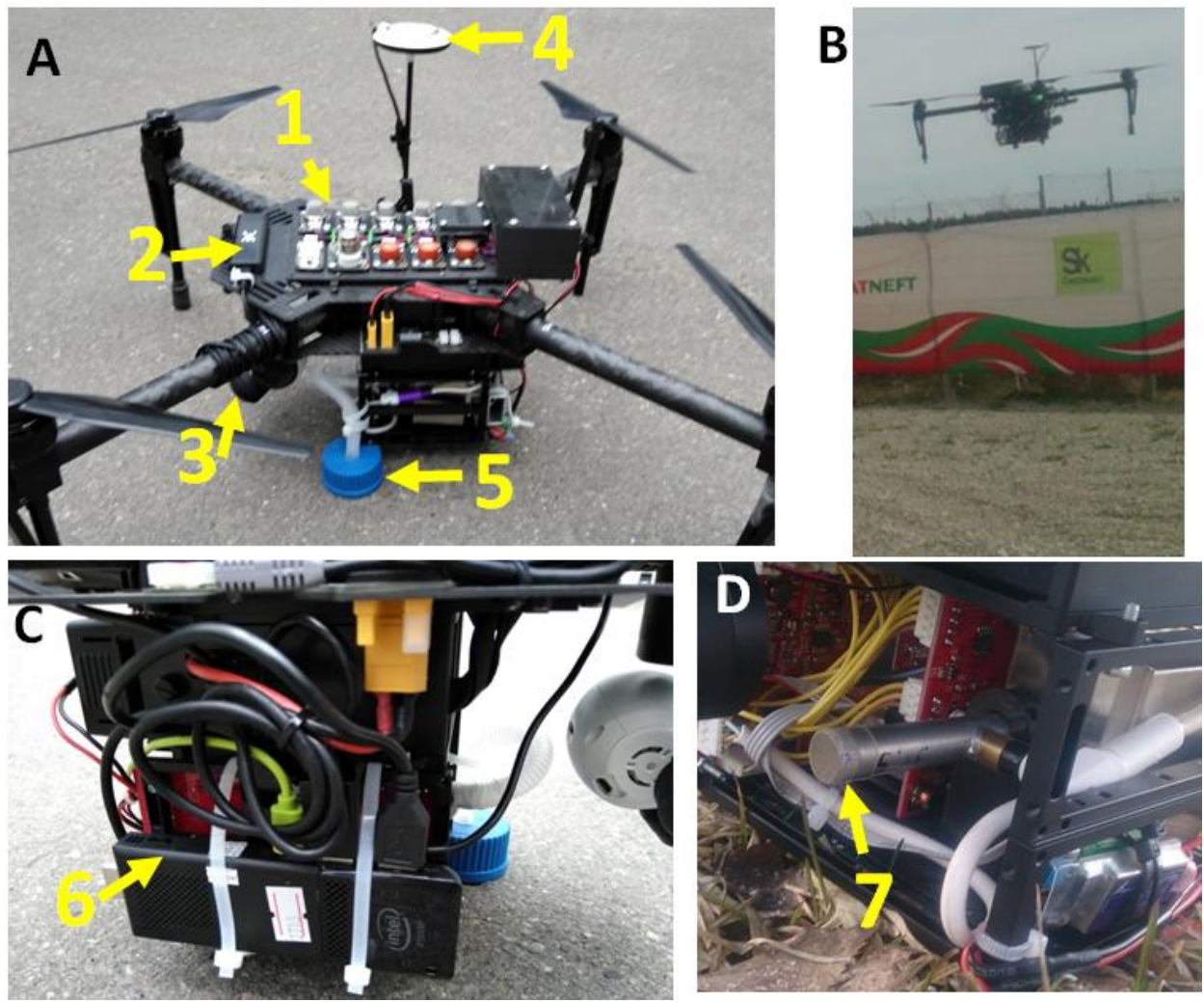

Fig.2 The developed measuring system. A - The photo of the assembled system: 1 – array of gas sensors, 2 – the 3G modem, 3 – the camera, 4 – GPS antenna, 5 – suction cup. B – the hovering measuring system. C – side view: 6 - the onboard computer,. D – the installed FAIMS spectrometer (7) without suction cups.

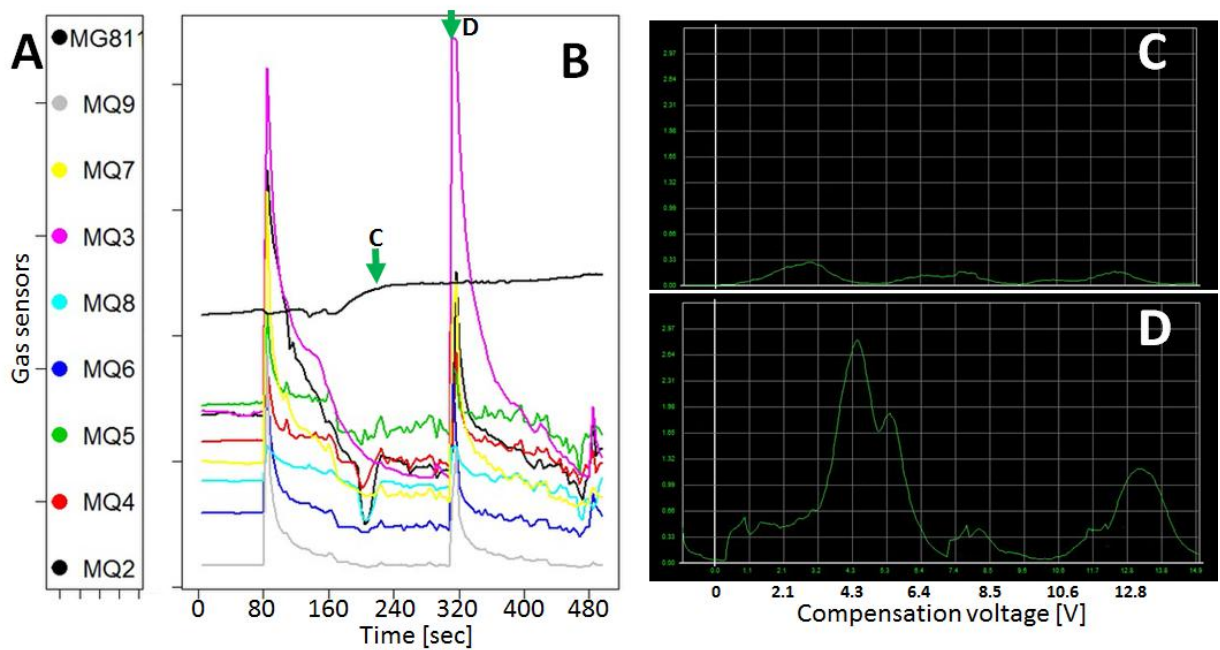

Fig.3. The signal obtained from the developed system during the flight. A- The list of the gas sensors, B – the dependence of the sensors output signal on time. C – the ionogram detected by FAIMS spectrometer when the quadrocopter is in the clean air. D - the ionogram detected by FAIMS spectrometer when the quadrocopter is in the cloud of odorous gas.

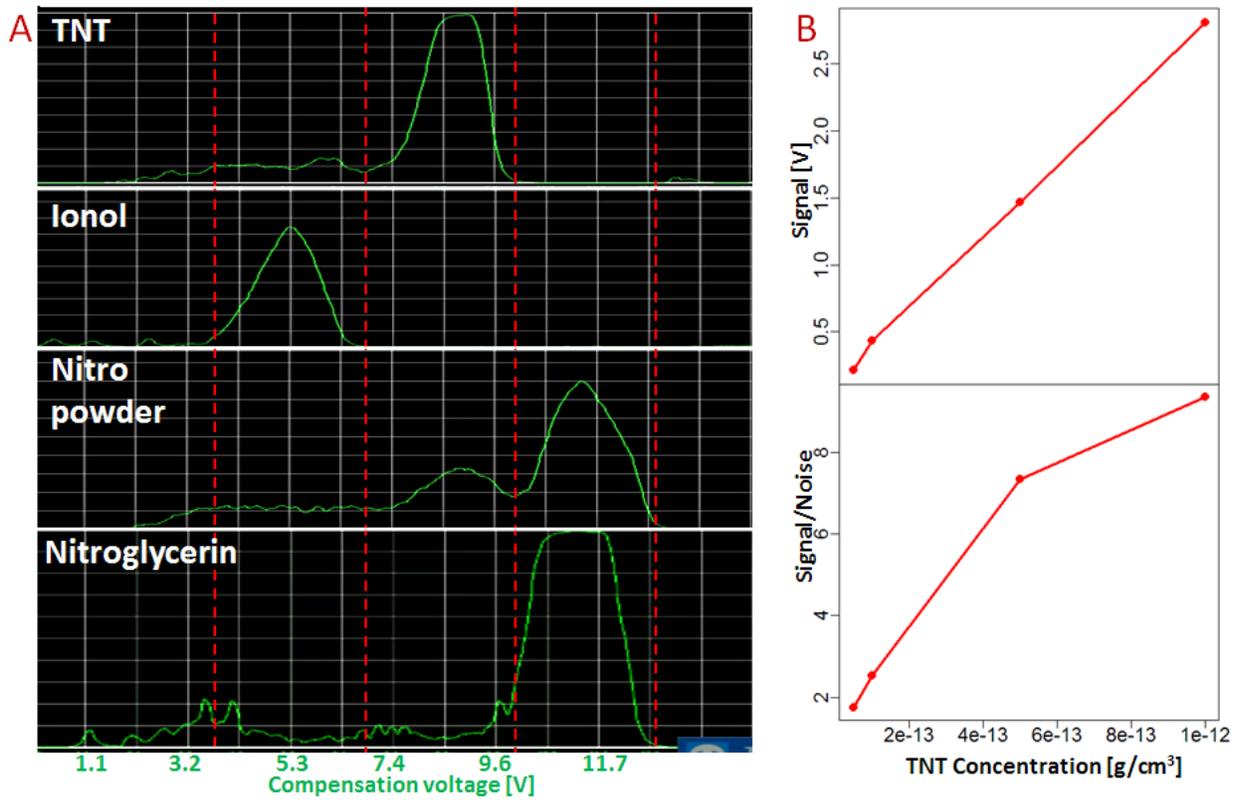

Fig.4 A - The ionogram corresponding to the traces of different chemicals: trinitrotoluene (TNT), Ionol, nitro powder and nitroglycerine. B – The value of signal and signal/noise ratio as a function of the TNT concentration in the air.

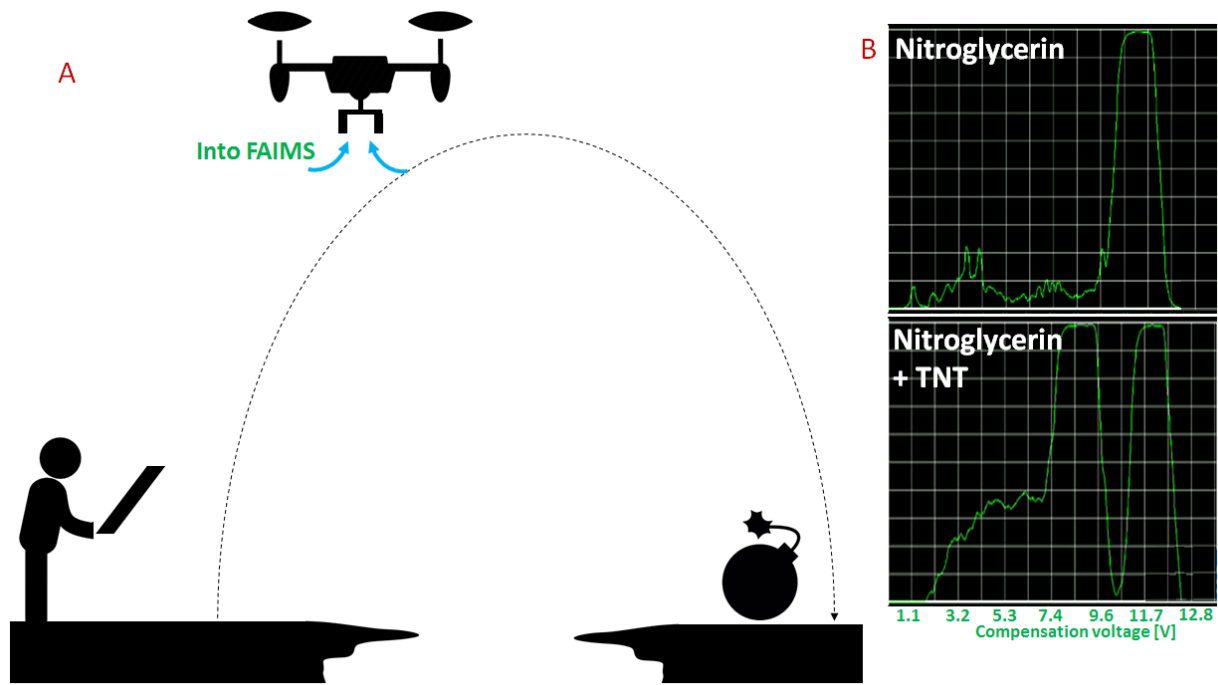

Fig.5 A - The schematic representation of the detection of explosives in the hard-to-reach places.

B – detection of nitroglycerine and simultaneous detection of nitroglycerine and TNT.